\documentclass{kluwer}    
\usepackage{graphicx}

\usepackage{floatflt}

\begin{document}
\begin{article}
\begin{opening}
\title{MHD Simulations of the ISM: The Importance of the Galactic Magnetic Field on the ISM ``Phases''}
\author{Miguel \surname{de Avillez}}
\runningauthor{M.A. de Avillez \& D. Breitschwerdt}
\runningtitle{The Importance of the Magnetic field on the ISM ``Phases''} \institute{Department of Mathematics, University of
\'Evora, R. Rom\~ao Ramalho 59, 7000 Evora, Portugal; Email:
\texttt{mavillez@galaxy.lca.uevora.pt}}
\author{Dieter \surname{Breitschwerdt}}
\institute{Max-Planck-Institut f\"ur Extraterrestrische Physik,
        Giessenbachstra{\ss}e, Postfach 1312, 85741 Garching, Germany; Email:
\texttt{breitsch@mpe.mpg.de}}
\date{September 29, 2003}

\begin{abstract}
  We have carried out 1.25 pc resolution MHD simulations of the ISM,
  on a Cartesian grid of $0 \leq (x,y) \leq 1$ kpc size in the
  galactic plane and $-10 \leq z \leq 10$ kpc into the halo, thus
  being able to fully trace the time-dependent evolution of the
  galactic fountain. The simulations show that large scale gas streams
  emerge, driven by SN explosions, which are responsible for the
  formation and destruction of shocked compressed layers. The shocked
  gas can have densities as high as 800 cm$^{-3}$ and lifetimes up to
  15 Myr. The cold gas is distributed into filaments which tend to
  show a preferred orientation due to the anisotropy of the flow
  induced by the galactic magnetic field. Ram pressure dominates the
  flow in the unstable branch $10^{2}<$T$\leq 10^{3.9}$ K, while for
  T$\leq 100$ K (stable branch) magnetic pressure takes over. Near
  supernovae thermal and ram pressures determine the dynamics of the
  flow. Up to $80\%$ of the mass in the disk is concentrated in the
  thermally unstable regime $10^{2}<$T$\leq 10^{3.9}$ K with
  $\sim30\%$ of the disk mass enclosed in the T$\leq 10^{3}$ K gas.
  The hot gas in contrast is controlled by the thermal pressure, since
  magnetic field lines are swept towards the dense compressed walls.

\end{abstract}

\end{opening}

\section{Introduction}
\label{intro}

Modelling the ISM with high spatial resolution allows us to tackle
a set of problems simultaneously, encompassing both large and
small scales, provided that the appropriate grid size, resolution
and numerical tools (e.g. adaptive mesh refinement) are used. To
name just the most important ones, global modelling yields
information on the formation and lifetimes of molecular clouds,
how star-forming regions are influenced by large-scale flows in
the ISM, and which dynamical r\^ole SNe and superbubbles play in
triggering local and global star formation. In this paper we
investigate the effects of the magnetic field on the dynamics and
evolution of the cold gas in the disk, the relative importance of
the field for the ISM ``phases'', and its influence on the support of
shocked compressed layers. 

\section{Model and Simulations}

In the current work we report on kpc-scale simulations of the ISM on a
cartesian grid of $0 \leq (x,y) \leq 1$ kpc size in the Galactic plane
and $-10 \leq z \leq 10$ kpc into the halo, using a modified version
of the 3D SN driven ISM model of Avillez (2000) coupled to a
three-dimensional MHD code that uses adaptive mesh refinement (AMR) in
a block-based structure in combination with Message Passing Interface
(MPI) suitable for massive parallel computations.

The modified model includes a fixed gravitational field provided by
the stars in the disk, radiative cooling assuming optically thin gas
in collisional ionization equilibrium, and uniform heating due to
starlight. The radiative cooling function is a tabulated version of
that shown in Figure~2 of Dalgarno \& McCray (1972) with an ionization
fraction of 0.1 at temperatures below $10^4$~K and a temperature
cutoff at 10~K.  Background heating due to starlight varies with $z$
as described in Wolfire et al.\ (1995); in the midplane, at $z=0$, it
is chosen to initially balance radiative cooling at 9000 K. With the
inclusion of background heating the gas becomes thermally stable at
T$\leq 100$ K and $10^{3.9}<$T$\leq 10^{4.2}$ K.

The prime sources of mass, momentum and energy are supernovae whose
setup is similar to that described in Avillez (2000) with some
modifications. In the present work SNe type Ia with a scale height of
325 pc are also included. In regions with a density and temperature
thresholds of 10 cm$^{-3}$ and 100 K, respectively, an initial mass
function is applied to determine the number of OB stars and their
masses, forming an OB association. In agreement with observations
60~\% of these stars explode within the association, while the
remaining lowest mass stars occur in the field. The location of the
field stars is determined kinematically by attributing to each star a
random direction and a velocity. The associations are allowed to form
in a layer with a scale height of 46 pc, while the field stars occur
in a layer with a scale height of 90 pc. The time interval between the
explosions of all OB stars is determined by their main sequence life
time.

The interstellar gas is initially setup with a density stratification
distribution that includes the cold, cool, warm, ionized and hot gas
``phases'' in the Galaxy as described by Ferri\`ere (1998). In these
simulations the Galactic SN rate has been used and the canonical
energy of explosion is $10^{51}$ erg for all types of SNe. At the
beginning of the simulations the uniform field components along the
three axes are given by $\vec{B}_{u}=(B_{u,0} (n(z)/n_0)^{1/2},0,0)$,
where $B_{u,0}=3~\mu$G is the field strength and $n(z)$ is the number
density of the gas as a function of distance from the Galactic
midplane and $n_0=1 \, {\rm cm}^{-3}$ is the average midplane density.
The random field component is set to zero in the beginning of the
simulations. This component is built up during the first millions of
years of evolution as a result of turbulent motions, mainly induced by
SN explosions. Self-gravity is not included in these calculations.

The computational grid has a resolution of 10 pc, except in the layer
between -500 and 500 pc, where three levels of AMR are used, yielding
a finest level resolution of 1.25 pc. Periodic boundary conditions are
applied along the four vertical boundary faces, while outflow boundary
conditions are imposed at the top ($z=10$ kpc) and bottom ($z=-10$
kpc) boundaries. All the simulations were evolved for 400 Myr
corresponding to several sound crossing times along the fountain as
required for the establishment of the duty disk-halo-disk cycle and
the global dynamical equilibrium.

\section{Results}
\subsection{Global Evolution}

The initial evolution of the magnetized disk
is similar to that seen in the HD runs (Avillez 2000, Avillez \&
Breitschwerdt 2003), that is, the initial stratified distribution
does not hold for long as a result of the lack of equilibrium between
gravity and (thermal, kinetic and turbulent) pressure during the
``switch-on phase'' of SN activity. As a consequence the gas in the
upper and bottom parts of the grid collapses into the midplane,
leaving low density material in its place. However, in the MHD run it
takes a longer time for the collapse to be completed as a result of
the magnetic pressure and tension forces. As soon as the system has
collapsed and enough supernovae have gone off in the disk building up
the required pressure support, transport into the halo is not
prevented, although the escape of the gas takes a few tens of Myr to
occur. The crucial point is that a huge thermal overpressure due to
combined supernova explosions can sweep the magnetic field into dense
filaments and punch holes into the extended warm and ionized H{\sc i}
layers.  Once such pressure release valves have been set up, there is
no way from keeping the hot over-pressured plasma to follow the
pressure gradient into the halo.  As a consequence the duty
disk-halo-disk cycle of the hot gas is fully established, which
combined to the input of energy into the ISM by SNe, diffuse heating,
energy lost by radiative cooling and magnetic pressure leads the
system to evolve into a dynamical equilibrium state within some 200
Myr (Kahn 1981).

\begin{figure}[t]
\centering
\vspace*{1.2in}
Left panel: mavillez$\_$fig1a, Right panel: mavillez$\_$fig1b
\vspace*{1.3in}
\caption{Density and magnetic field distribution in the Galactic midplane after 374 Myr of disk evolution. The resolution of the finest AMR level is 1.25 pc.}
\label{fig1}
\end{figure}

Fig.~\ref{fig1} shows slices of the three-dimensional data cube of
the density and magnetic field distributions in the Galactic midplane.
The highest density ($n\ge 100$ cm$^{-3}$) gas tends to be confined
into shocked compressed layers that form in regions where several
large scale streams of convergent flow (associated with laminar flows
driven by SNe) occur. The compressed regions are filamentary in
structure, tend to be aligned with the local field and are associated
with the highest field strengths\footnote{A similar result is seen in
  similar HD simulations. However, there is no preferable orientation
  for the filaments.}. The formation time of these high density
structures depends on how much mass is carried by the convergent
flows, how strong compression is and on the rate of cooling of the
compressed region. The compressed layers have on average lifetimes of
10-15 Myr. The streams that promote the formation of the shocked dense
layers are also responsible for their destruction. The filamentary
structures are thicker in these simulations than in similar HD
simulations, because the compression of a shock wave at given Mach
number must go into increase of thermal and magnetic pressure in the
MHD case and only into thermal in HD.

\begin{figure}[t]
\centering
\vspace*{1.1in}
Left panel: mavillez$\_$fig2a, Right panel: mavillez$\_$fig2b
\vspace*{1.1in}
\caption{Scatter plot of the magnetic field strength as a function
  of density for the T$\leq 10^{3}$ (red) and $10^{4}<$T$\leq10^{5.5}$ K (black) regimes (left panel) and $10^{3}<$T$\leq 10^{4}$ (black) and T$>10^{5.5}$ K (red) regimes (right panel) at 400 Myr of evolution. }
\label{scatterDB}
\end{figure}

\subsection{Field Dependence with Density}

The simulations started with a fixed value for the magnetic field
varying with $z$ and having a midplane value of $3\times 10^{-6}$ G.
During the evolution of the system thermal and dynamical processes
broaden the distribution of the field strength in such a way that
after the global dynamical equilibrium has been set up the field
strength in the disk spans two orders of magnitude from $10^{-7}$ to
$10^{-5}$ G as can be seen in Fig.~\ref{scatterDB}. The figure shows a
scatter plot of the field strength as function of density in the
simulated disk after 400 Myr of evolution.

Both panels in the figure show a large scatter in the field for the
same density. This lack of correlation suggests that the field may not
follow the ``classical'' relation of $B\sim\rho^{1/2}$. This essentially
means that the equation of state is considerably softer than in the
pure magnetic case and hence the effective adiabatic index $\gamma$
must be well below unity.  This may be attributed to the variety of
processes that affect the energy balance of the magnetized plasma,
such as turbulence, heating and cooling.  Specifically the equation of
state becomes stiffer, if the dependence of the heating rate on
density is steeper than that of the cooling rate.

The large scatter in the field seen in Fig.~\ref{scatterDB} also
suggests that the field is being driven by the inertial motions,
rather than it being the agent determining the motions. In the latter
case the field would not be strongly distorted, and it would direct
the motions predominantly along the field lines.  The high field
variability is also seen in the right panel of Fig.~1, which shows a
highly turbulent field, that seems to be uncorrelated with the
density.

\subsection{Pressure Variation with Temperature}

Left panel of Fig.~\ref{scatter1} shows the scatter points of the
thermal (black) and magnetic (red) pressures as a function of
temperature, while the right panel shows the average magnetic, thermal
and ram pressures as function of temperature. From the figure it is
seen that all the T$\leq 10^{2}$ K gas has $P_{B}>P_{ram}\gg P_{th}$,
demonstrating that magnetically dominated regions do exist, while the
hot gas has the highest thermal pressure and the lowest magnetic
pressure. For T$\leq 10^{2}$ the relation $B \propto \rho^{\alpha}$ does
not hold, as $\alpha$ varies between -0.06 and 0.085 (Avillez \&
Breitschwerdt 2004) indicating that the magnetic and thermal pressures
are independent. A result that is consistent with the almost zero
variation of magnetic pressure with temperature seen in the right
panel of Fig.~\ref{scatter1}.

\begin{figure}[t]
\centering
\vspace*{1.1in}
Left panel: mavillez$\_$fig3a, Right panel: mavillez$\_$fig3b
\vspace*{1.1in}
\caption{Left panel shows a scatter plot of the magnetic (red) and thermal (black) pressures as functions of temperature at 400 Myr. Right panel shows the average magnetic (red), thermal (black) and ram (green) pressures as functions of temperature at 400 Myr. This plot is obtained by calculating the average pressure in a certain temperature bin. The red and black curves represent the average values of the pressures seen in the left panel.}
 \label{scatter1}
\end{figure}

The figure also shows that for $10^{2}<$T$<10^{6}$ K ram pressure
dominates over magnetic and thermal pressures, that is, ram pressure
determines the dynamics of the flow, and therefore, the magnetic
pressure does not act as a restoring force as it was already suggested
by the lack of correlation between the field strength and the density
(for a detailed discussion see Passot \& V\'azquez-Semadeni 2003).
The alternating dominance between magnetic and ram pressures seen for
$10^{6}<$T$<3\times 10^{6}$ K results from the fact that about half of
the explosion energy in supernovae becomes kinetic, and therefore, ram
pressure becomes also important near the energy sources.

The dynamical picture that emerges from these simulations is that
thermal pressure gradients dominate mostly in the neighborhood of
supernovae, which drive motions whose ram pressures are dominant over
the mean thermal pressure (away from the energy sources) and the
magnetic pressure. The magnetic field is dynamically
important at low temperatures. The field does not act only by its
pressure gradient, but it acts \emph{dynamically} also by magnetic
tension forces, $1/4\pi (\vec{B}\nabla) \vec{B}$. Whereas magnetic
pressure is isotropic, magnetic tension is always \emph{along} the
lines of force. Equilibrium only exists in force-free configurations
for which $\nabla \times \vec{B} = 0$. However, if there is a thermal
pressure then it is practically impossible to balance the magnetic
stresses and most of the configurations will be unstable.  In other
words, in a dynamical system like the ISM, which includes systematic
gas motions, $\vec{j} \times \vec{B}$-forces are inevitable, and hence
the magnetic field can be dominant. Thus, regions with a
high field strength such as the cold clouds should be magnetically
controlled.
\begin{figure*}[t]
  \centering
\vspace*{1.1in}
Left panel: mavillez$\_$fig4a, Right panel: mavillez$\_$fig4b
\vspace*{1.1in}

\caption{Averaged mass weighted (left panel) and cumulative (right panel)
histograms of the density in the Galactic disk calculated between 300 and 400 Myr using 101 snapshots with 1 Myr of interval. The black and blue solid lines represent gas in the thermally stable regimes at T$\leq 10^{2}$ K and $10^{3.9}<$T$\leq 10^{4.2}$ K, the orange solid and dashed lines represent the thermally unstable gas with $10^{2}<$T$\leq 10^{3}$ K and $10^{3}<$T$\leq 10^{3.9}$ K, respectively. The gas with $10^{4.2}<$T$\leq 10^{5.5}$ K is represented by the green line, while the hot (T$>10^{5.5}$ K) gas is represented by the red line.
}
\label{mass}
\end{figure*}

\subsection{Distribution of the ISM Mass}

Up to $80\%$ of the ISM mass in the simulated disk is concentrated in
the unstable regime at $10^{2}<$T$\leq 10^{3.9}$ K (orange curves in
Fig.~\ref{mass}) where $\sim 30\%$ of the ISM mass is concentrated in
$10^{2}<$T$ <10^{3}$ K gas (dashed orange curve), while the
$10^{3}<$T$ \leq 10^{3.9}$ K gas (solid orange line) encloses some
$50\%$ of the mass. The stable regimes T$\leq 10^{2}$ K and
$10^{3.9}<$T$\leq 10^{4.2}$ K enclose $\sim 1\%$ and $\sim 10\%$ of
the disk mass, respectively. Note that most of the cold gas is
concentrated in the unstable branch at temperatures T$\leq 10^{3}$ K
and has densities up to 100 cm$^{-3}$ while gas with highest density
has T$>10^{3}$ K. The hot gas amounts to less than $1\%$, while the
unstable regime with $10^{4.2}<$T$\leq 10^{5.5}$ encloses $\sim 10\%$
of the total mass.

\section{Summary}

The highest resolution MHD simulation of the ISM carried out to date,
discussed in this paper, shows that large scale streams driven by SNe
are responsible for the formation and destruction of high density
clouds in shocked compressed layers. The clouds formed within the
shocked gas can have densities as high as 800 cm$^{-3}$ and lifetimes
up to 15 Myr. Ram pressure dominates most of the coolest flows (in the
unstable branch $10^{2}<$T$\leq 10^{3.9}$ K) except at temperatures
below 100 K (in the stable branch), where magnetic pressure takes
over. 

Most of the mass in the disk is concentrated in the thermally unstable
regime for T$\leq 10^{3.9}$ K with some 30\% of the mass enclosed in
gas with T$\leq 10^{3}$ K. This is consistent with observations and
numerical simulations discussed by Heiles (2001), Gazol et al. (2001)
and Kritsuk \& Norman (2002).  However, these results are certainly in
disagreement with classical ISM theories (McKee \& Ostriker 1977), as
from our simulations up to $90\%$ of the ISM mass is found to be in
the \emph{thermally unstable} regimes with $10^{2}<$T$\leq 10^{3.9}$ K
($\sim 80\%$) and $10^{4.2} <$T$ \leq 10^{5.5}$ K ($\sim10\%$), where
it should not exist, if it were not for the importance of dynamical
processes. This calls into question the standard paradigm of an ISM
distributed over three phases and being in pressure equilibrium.

\acknowledgements We acknowledge fruitful discussions with the
referee, E. V\'azquez-Semadeni, and M.-M. Mac Low. MAA thanks A.I.
G\'omez de Castro for the financial support.

\end{article}

\begin{thebibliography}{}
  
\bibitem{avi00} Avillez, M.A. 2000, MNRAS, 315, 479
\bibitem{avi03a} Avillez, M.A., Breitschwerdt, D. 2003, A\&A (in press)
\bibitem{avi04} Avillez, M.A., Breitschwerdt, D. 2004, A\&A (submitted)
\bibitem{da72} Dalgarno, A., McCray, R.A. 1972, ARA\&A 10, 375
\bibitem{fe98} Ferri\`{e}re, K.M. 1998, ApJ 503, 700
\bibitem{ga01} Gazol, A., V\'azquez-Semadeni, E., Sanchez-Salcedo, J., Scalo, J. 2001, ApJ, 557, 121
\bibitem{he01} Heiles, C. 2001, ApJ, 551, L105
\bibitem{ka81} Kahn, F.D. 1981, in: ``Investigating the Universe'', ed.\ F.D.\ Kahn, Reidel Dordrecht, p.~1
\bibitem{kn02} Kritsuk, A., Norman, M. L. 2002, ApJ, 569, L127
\bibitem{mo77} McKee, C. F., Ostriker, J. P. 1977, ApJ 218, 148  %
\bibitem{pvs03} Passot, T., V\'azquez-Semadeni, E. 2003, A\&A, 398, 845
\bibitem{wo95} Wolfire, M.G., McKee, C.F., Hollenbach, D., Tielens, A.G.G.M., Bakes, E.L.O. 1995, ApJ, 443, 152


\end{thebibliography}
\end{document}